\documentclass[a4paper,11pt]{article}
\usepackage{pos}

\title{High-Energy Extractions from Horizonless Compact Objects}

\author*[a]{Parth Bambhaniya}
\author[a]{Elisabete M. de Gouveia Dal Pino}

\affiliation[a]{Instituto de Astronomia, Geofísica e Ciências Atmosféricas, Universidade de  São Paulo, IAG,\\
Rua do Matão, 1225, CEP: 05508-090 São Paulo - SP - Brazil}

\emailAdd{parth.bambhaniya@usp.br}
\emailAdd{dalpino@iag.usp.br}

\abstract{High-energy astrophysical sources such as active galactic nuclei, quasars, X-ray binaries, and gamma-ray bursts are powered by mechanisms that convert gravitational or rotational energy into radiation, jets, and relativistic outflows. Understanding the physics of these processes remains a major challenge. Black holes have traditionally served as the central engines behind such phenomena, with well-established energy extraction mechanisms including the Penrose process, the Blandford–Znajek process, and the Banados–Silk–West mechanism. However, studies in general relativity indicate that, under certain conditions, gravitational collapse may lead to the formation of naked singularities or other horizonless compact objects, which could in principle allow more efficient energy extraction than classical black holes. This brief review summarizes recent progress on energy extraction mechanisms in naked singularity spacetimes. We examine the roles of rotation, electromagnetic fields, and particle interactions in shaping extraction efficiency and dynamics. Particular attention is given to negative energy orbits and ergoregion physics, which enable Penrose type and magnetic Penrose mechanisms without an event horizon. We also discuss collisional Penrose processes and particle acceleration near the singularity, emphasizing their potential astrophysical implications. By comparing extraction efficiencies and physical conditions in black holes and naked singularities, we highlight how the absence of a horizon fundamentally alters the dynamics of energy release. These results suggest that naked singularities may serve as natural laboratories for strong field gravity and as alternative engines for high-energy astrophysical phenomena in the era of multi-messenger observations.}

\FullConference{
High Energy Phenomena in Relativistic Outflows IX (HEPRO-IX)\\
4-8 August 2025\\
Rio de Janeiro, Brazil\\}

\begin{document}
\maketitle

\section{Introduction}

Compact objects such as active galactic nuclei, quasars, and X-ray binaries exhibit extraordinary luminosities, suggesting efficient conversion of gravitational and rotational energy into radiation and relativistic outflows. Explaining how this transformation occurs remains a central problem in high-energy astrophysics. The Penrose process was the first relativistic mechanism to demonstrate that rotational energy could be extracted from a Kerr black hole through particle fission in the ergoregion, where negative energy orbits exist \cite{Penrose:1971uk, Patel:2022jbk}. Later developments introduced magnetic and collisional extensions of the Penrose process \cite{Wagh:1985vuj,Tursunov:2019oiq}, where magnetic fields or particle collisions substantially enhance extraction efficiency. Magnetohydrodynamic frameworks such as the Blandford–Znajek (BZ) process \cite{Bz} describe how magnetic field lines anchored to a rotating compact object extract its rotational energy as Poynting flux, launching relativistic jets. The Blandford–Payne model complements this picture by demonstrating how plasma from the accretion disk can be centrifugally accelerated along magnetic field lines \cite{BP}. These processes form the theoretical backbone for interpreting observed jet structures and variability in compact astrophysical systems.
Although such mechanisms were primarily formulated for black holes, they can operate in horizonless geometries such as naked singularities. The absence of an event horizon allows energy extracted near the singular region to escape freely, providing a natural testbed for studying high-efficiency astrophysical engines. The Janis–Newman–Winicour (JNW) naked singularity in particular offers a valuable model for exploring these ideas \cite{Patel:2023efv}.

\section{Energy Extraction Mechanisms from Compact Objects}

\subsection{Penrose Process and Magnetic Extensions}
The Penrose process \cite{Penrose:1971uk} involves a particle entering the ergoregion and splitting into two fragments, one carrying negative energy and falling inward while the other escapes with excess energy. This allows extraction of rotational energy through the conservation of angular momentum. The magnetic Penrose process \cite{Wagh:1985vuj} enhances this scenario by introducing electromagnetic interactions. Charged particles in magnetic fields can access broader ranges of negative energy states, greatly increasing efficiency. In certain configurations, the extracted energy can exceed the input energy, yielding efficiencies above one hundred percent, making this mechanism especially relevant for high-energy astrophysical jets.

\subsection{Blandford–Znajek and MHD Mechanisms}
The Blandford–Znajek (BZ) process \cite{Bz} represents a large-scale magnetohydrodynamic energy extraction mechanism. Here, magnetic field lines penetrating the rotating compact object act as conduits for electromagnetic energy, launching Poynting flux dominated outflows. The process operates without requiring particle disintegration, relying instead on the twisting of field lines by frame dragging. The Blandford–Payne mechanism adds a complementary disk-driven component, showing how accreting plasma can be flung out along magnetic lines anchored in the disk \cite{BP}. Together, these mechanisms explain the persistent and collimated jets observed in AGN, microquasars, and gamma-ray bursts.

\subsection{Collisional and Super-Penrose Processes}
High-energy particle collisions near compact objects provide another route for energy extraction. The Banados–Silk–West (BSW) mechanism \cite{Banados:2009pr} reveals that collisions close to the horizon of extremal Kerr black holes can achieve extremely high center-of-mass energies. In horizonless geometries, this efficiency can increase dramatically, giving rise to the super-Penrose process \cite{Zaslavskii:2022nbm}, where outgoing fragments achieve unbounded energies. Such collisions illustrate the potential of compact objects to act as natural particle accelerators operating at energies far beyond those accessible in terrestrial experiments.

\section{Energy Extraction from Naked Singularities and Negative Energy Orbits}
Naked singularities are theoretical solutions of general relativity where the region near the singularity  remains visible to external observers. The lack of an event horizon allows energetic particles and radiation to escape, making these configurations promising candidates for efficient astrophysical power sources. They also provide an arena to test cosmic censorship and explore how gravitational, electromagnetic, and plasma processes behave under extreme curvature. In rotating geometries, negative energy orbits enable rotational energy extraction. The rotating JNW spacetime, sourced by a scalar field, lacks a traditional ergoregion \cite{Karmakar:2017lho}, yet retains strong frame-dragging effects that allow negative energy trajectories. Analyses show that even without a horizon, particles can gain large amounts of energy through collisions and escape to infinity \cite{Patel:2023efv}. The scalar charge modifies the effective potential, influencing both orbit stability and extraction efficiency. This dependence on scalar field strength suggests that deviations from the Kerr metric may enhance or suppress energy extraction depending on the degree of compactness. Hence, rotating naked singularities serve as exceptional laboratories for exploring non-Kerr spacetimes within the general relativity.

\subsection{Magnetic Coupling and Enhanced Efficiency}

Magnetic fields amplify the complexity of particle motion around naked singularities. Even weak magnetic fields can significantly alter particle trajectories by coupling electromagnetic and gravitational effects. In magnetized JNW configurations, charged particles extract energy both from rotation and magnetic interaction, leading to increased overall efficiency \cite{Patel:2023efv}. Synchrotron radiation and magnetic reconnection can further convert magnetic energy into radiation \cite{deGouveiaDalPino:2003mu}, intensifying high-energy emission. The absence of a horizon ensures that this energy is not reabsorbed, allowing the full luminosity to reach distant observers. Such dynamics may explain rapid variability and intense flaring observed in compact sources, linking magnetic topology to observable signatures.

\subsection{Charged Naked Singularities and Extreme Collisions}

Overcharged Reissner–Nordström naked singularities provide another setting for studying unbounded energy extraction. When the charge-to-mass ratio exceeds unity, strong electromagnetic repulsion and gravitational attraction create regions where particle collisions yield enormous center-of-mass energies \cite{Zaslavskii:2022nbm}. These systems could act as astrophysical accelerators, producing particles with energies high enough to explain ultra-high-energy cosmic rays and transient bursts, which are difficult to model with black hole-based mechanisms.

\section{Discussion and Conclusions}

Energy extraction from naked singularities may underpin several high-energy phenomena. The combination of strong curvature, frame dragging, and magnetic coupling can generate distinctive spectral, timing, and polarization features. These include rapid variability, strong linear polarization, and enhanced jet power compared to black holes of similar mass. Such effects could be probed through upcoming high-resolution and polarimetric observations by the Event Horizon Telescope and the Cherenkov Telescope Array Observatory. Detecting deviations from Kerr-like emission patterns would offer critical evidence for horizonless compact objects and provide insights into gravitational collapse beyond the traditional black hole scenario. Naked singularities offer a unique platform for examining energy extraction in strong gravitational and electromagnetic fields. Their horizonless nature allows unrestricted emission, enabling higher extraction efficiencies than black holes. Studies of rotating and magnetized JNW spacetimes demonstrate that energy extraction can occur without a classical ergoregion \cite{Patel:2023efv}. Magnetic coupling plays a central role, while charged naked singularities can realize unbounded energy gains through the super-Penrose process \cite{Zaslavskii:2022nbm}. 
These results suggest that naked singularities could act as powerful astrophysical engines and natural cosmic accelerators. Distinguishing them observationally requires precise multiwavelength, timing, and polarimetric data. The theoretical framework established in \cite{Patel:2023efv} provides a foundation for interpreting future observations and deepening our understanding of the connection between gravity, magnetism, and high-energy radiation in horizonless compact objects.

\acknowledgments{\noindent P. Bambhaniya and Elisabete M. de Gouveia Dal Pino acknowledge support from the São Paulo State Funding Agency FAPESP (grant 2024/09383-4).}


\begin{thebibliography}{99}
\bibitem{Penrose:1971uk}
R.~Penrose and R.~M.~Floyd,
\href{https://www.nature.com/articles/physci229177a0}{Nature \textbf{229}, 177-179, (1971).}

\bibitem{Patel:2022jbk}
V.~Patel, K.~Acharya, P.~Bambhaniya and P.~S.~Joshi,
\href{https://www.mdpi.com/2218-1997/8/11/571}{Universe \textbf{8}, 571, (2022).}

\bibitem{Wagh:1985vuj}
S.~M.~Wagh, S.~V.~Dhurandhar and N.~Dadhich,
\href{https://ui.adsabs.harvard.edu/abs/1985ApJ...290...12W/abstract}{Atrophys. J. \textbf{301}, 1018, (1986).}


\bibitem{Tursunov:2019oiq}
A.~Tursunov and N.~Dadhich,
\href{https://www.mdpi.com/2218-1997/5/5/125}{Universe \textbf{5}, no.5, 125 (2019).}


\bibitem{Bz}
R. D. Blandford and R. L. Znajek,
\href{https://articles.adsabs.harvard.edu/pdf/1977MNRAS.179..433B}{Mon. Not. R. Astron. Soc. {\bf 179}, 433 (1976).}

\bibitem{BP}
R. D. Blandford, D. G. Payne, 
\href{https://doi.org/10.1093/mnras/199.4.883}{Mon. Not. R. Astron. Soc. {\bf 199}, 883-903 (1982).}

\bibitem{Banados:2009pr}
M.~Banados, J.~Silk and S.~M.~West,
\href{https://journals.aps.org/prl/abstract/10.1103/PhysRevLett.103.111102}{Phys. Rev. Lett. \textbf{103}, 111102, (2009).}

\bibitem{deGouveiaDalPino:2003mu}
E.~M.~de Gouveia Dal Pino and A.~Lazarian,
\href{https://www.aanda.org/articles/aa/abs/2005/39/aa2590-04/aa2590-04.html}{A\&A 441, 845–853 (2005).}

\bibitem{Patel:2023efv}
V.~Patel, K.~Acharya, P.~Bambhaniya and P.~S.~Joshi,
\href{https://doi.org/10.1103/PhysRevD.107.064036}{Phys. Rev. D \textbf{107}, 064036 (2023).}

\bibitem{Karmakar:2017lho}
T.~Karmakar and T.~Sarkar,
\href{https://link.springer.com/article/10.1007/s10714-018-2408-y}{Gen. Rel. Grav. \textbf{50}, 85, (2018).}

\bibitem{Zaslavskii:2022nbm}
O.~B.~Zaslavskii,
\href{https://journals.aps.org/prd/abstract/10.1103/PhysRevD.105.124043}{Phys. Rev. D \textbf{105}, 124043, (2022).}





\end{thebibliography}
\end{document}